\documentclass[twoside,11pt] {article}

\usepackage{graphicx}
\usepackage{amssymb}
\usepackage{amsmath}

\begin{document}

\title{Dark matter and dark energy production in quantum model of the universe}
\author{V.E. Kuzmichev and V.V. Kuzmichev\\[0.5cm]
\itshape Bogolyubov Institute for Theoretical Physics,\\
\itshape Nat. Acad. of Sci. of Ukraine, Kiev, 03143 Ukraine}
\date{}

\maketitle

\pagestyle{myheadings} \thispagestyle{plain} \markboth{V.E.
Kuzmichev and V.V. Kuzmichev}{Dark matter and dark energy
production} \setcounter{page}{1}

\begin{abstract}
The matter/energy structure of the homogeneous, isotropic, and
spatially closed universe is studied. The quantum model under
consideration predicts an existence of two types of collective
quantum states in the universe. The states of one type
characterize a gravitational field, the others describe a matter
(uniform scalar) field. In the first stage of the evolution of the
universe a primordial scalar field evolves slowly into its
vacuum-like state. From the point of view of semiclassical
description the early universe is filled with primordial radiation
and is charge symmetric in this stage. In the second stage the
scalar field oscillates about an equilibrium due to the quantum
fluctuations. The universe is being filled with matter in the form
of elementary quantum excitations of the vibrations of the scalar
field. The separate quantum excitations are characterized by
non-zero values of their energies (masses). Under the action of
gravitational forces mainly these excitations decay into ordinary
particles (baryons and leptons) and dark matter. The elementary
quantum excitations of the vibrations of the scalar field which
have not decayed up to now form dark energy. The numerical
estimations lead to realistic values of both the matter density
$\Omega_{M} \simeq 0.29$ (with the contributions from dark matter,
$\Omega_{DM} \simeq 0.25$, and optically bright baryons,
$\Omega_{stars} \simeq 0.0025$) and the dark energy density
$\Omega_{X} \simeq 0.71$ if one takes that the mean energy $\sim
10$ GeV is released in decay of dark energy quantum and fixes
baryonic component $\Omega_{B} = 0.04$ according to observational
data. The energy (mass) of dark energy quantum is equal to $\sim
17$ GeV and the energy $\gtrsim 2 \times 10^{10}$ GeV is needed in
order to detect it. Dark matter particle has the mass $\sim 6$
GeV. The Jeans mass for dark matter which is considered as a gas
of such massive particles is equal to $M_{J} \sim 10^{5}
M_{\odot}$.
\end{abstract}

\section{Introduction}
\label{Intro}

Observations indicate that overwhelming majority (about 96\%) of
matter/energy in the universe is in unknown form (see e.g.
Refs.~\cite{Ol,Sa} for reviews). The observed mass of stars gives
the value $\Omega_{stars} \simeq 0.005$ \cite{Co} or even smaller
$\Omega_{stars} \simeq 0.003^{+0.001}_{-0.002}$ \cite{PS} for the
density of visible (optically bright) baryons. Observations of the
cosmic microwave background radiation (CMB) and abundances of the
light elements in the universe suggest that the total density of
baryons is equal to $\Omega_{B} \simeq 0.04$ \cite{Ol,Sa,FS}. This
value is one order greater than $\Omega_{stars}$. It means that
most of baryonic matter today is not contained in stars and is
invisible (dark).

The CMB anisotropy measurements allow to determine the total
energy density $\Omega_{tot}$ and the matter component
$\Omega_{M}$. The recent data give the strong evidence that the
present-day universe is spatially flat (or very close to it) with
$\Omega_{tot} \simeq 1$ \cite{BNPS} and the mean matter density
equals $\Omega_{M} \simeq 0.3$ \cite{Ol}. The independent
information about $\Omega_{M}$ extracted from the high redshift
supernovae Ia data on the assumption that $\Omega_{tot} = 1$ gives
the close values: $\Omega_{M} = 0.28 \pm 0.05$ \cite{To} or
$\Omega_{M} = 0.29^{+0.05}_{-0.03}$ \cite{Ri}. The discrepancies
between $\Omega_{M}$ and $\Omega_{B}$ on the one hand and
$\Omega_{tot}$ and $\Omega_{M}$ on the other hand are signs that
there must exist non-baryonic dark matter with the density
$\Omega_{DM} = \Omega_{M} - \Omega_{B} \sim 0.3$ and some
mysterious cosmic substance (so-called dark energy \cite{OSB})
with the density $\Omega_{X} = \Omega_{tot} - \Omega_{M} \sim
0.7$. The origin and composition of both dark matter and dark
energy are unknown. Dark matter manifests itself in the universe
through the gravitational interaction. Its presence allows to
explain rotation curves for galaxies and large-scale structure of
the universe in the models with standard assumption of adiabatic
density perturbations \cite{Ol,Sa,Do}. Candidates for dark matter
and dark energy are discussed e.g. in reviews \cite{Ol,Sa,Do,PR}.
As regards dark energy it is worth mentioning that its expected
properties are unusual. It is unobservable (in no way could it be
detected in galaxies) and spatially homogeneous.

Thus the present data of modern cosmology pose the principle
question about the nature of the mass-energy constituents of the
universe and their percentage in the total energy density. Efforts
in this direction were focused on a choice of candidates for dark
matter and dark energy between known (real or hypothetical)
particles and fields. It is obvious that reasonable cosmological
theory must first of all answer the question why the densities
$\Omega_{M}$ and $\Omega_{X}$ in the present era are comparable
between themselves (so-called coincidence problem) and explain the
observed ratio $\Omega_{B}/\Omega_{stars} \sim O(10)$.

In the present paper the problem of dark matter and dark energy is
studied in the context of the quantum model of the universe
proposed in Refs.~\cite{K,KK}. The first attempts to give an
answer to the question about the nature of dark matter and dark
energy on the basis of quantum approach to cosmological problems
were made in Refs.~\cite{KK2,KK3}.

In Sec. 2 the basic equations of the quantum model of the
homogeneous, isotropic and spatially closed universe are given. It
is supposed that the universe is filled with primordial matter in
the form of the uniform scalar field. Time is introduced as an
embedding variable which describes a motion of some source. From
the point of view of semiclassical approach this source has a form
of radiation. The evolution of the universe can be conventionally
divided into two stages. At the first stage (Sec.~3) the scalar
field determines the vacuum energy density which slowly evolves
into the state with minimal density (vacuum-like state). Radiation
is present as a primordial source and universe is charge
symmetric. In Sec.~4 the second stage of the evolution of the
universe in considered. At this stage the scalar field oscillates
about the equilibrium vacuum-like state and the universe is being
filled with matter in the form of elementary quantum excitations
of the vibrations of the scalar field. These excitations have the
non-zero energies (masses). The wavefunction is an amplitude of
the probability wave of the universe to be in the state with given
values of two quantum numbers. One quantum number characterizes
the gravitational field, while another relates to the scalar
field. In Sec.~5 the simple model of creation of matter in the
ordinary forms as a result of decay of elementary quantum
excitations of the vibrations of the scalar field under the action
of gravitational forces is proposed. The numerical estimations of
the percentage of dark matter and dark energy in the present-day
universe are given. A comparison of theoretical calculations with
integrated data from WMAP, other CMB experiments, HST key project
and supernovae observations \cite{Sp} is made. In Sec.~6 some
conclusions are drawn.

\section{Basic equations of the model}
\label{Basic}

Let us consider the quantum model of the homogeneous, isotropic
and spatially closed universe filled with primordial matter in the
form of the uniform scalar field $\phi$ with some potential energy
density $V(\phi)$. The time-dependent equation which describes
such a universe has a form \cite{K,KK} (here and below we use
dimensionless variables where the length is measured in units of
$l_{P} = \sqrt{2G/(3\pi)}$ and the energy density in $\rho_{P} =
3/(8 \pi G l_{P}^{2})$)
\begin{equation}\label{1}
    i\,\partial_{\mathcal{T}} \Psi = \hat{\mathcal{H}} \Psi,
\end{equation}
where
\begin{equation}\label{2}
    \hat{\mathcal{H}} = \frac{1}{2} \left(\partial_{a}^{2} -
    \frac{2}{a^{2}}\,\partial_{\phi}^{2} - a^{2} + a^{4} V(\phi)
    \right )
\end{equation}
is a Hamiltonian-like operator. The wavefunction $\Psi$ depends on
the cosmological scale factor $a$, scalar field $\phi$, and time
coordinate $\mathcal{T}$. In derivation of Eq.~(\ref{1}) time
$\mathcal{T}$ is introduced as an additional (embedding) variable
which describes a motion of a source in a form of relativistic
matter of an arbitrary nature from the point of view of
semiclassical approach. It is related to the synchronous proper
time $t$ by the differential equation: $dt = a\, d\mathcal{T}$
\cite{K}. Eq.~(\ref{1}) allows a particular solution with
separable variables
\begin{equation}\label{3}
    \Psi = \mbox{e}^{\frac{i}{2} E \mathcal{T}} \psi_{E},
\end{equation}
where the function $\psi_{E}$ is given in $(a,\phi)$-space of two
variables and satisfies the time-independent equation
\begin{equation}\label{4}
 \left( -\,\partial _{a}^{2} +  a^{2} - a^{4} \hat{\rho}_{\phi}  - E  \right)
 \psi _{E} = 0.
\end{equation}
Here the operator
\begin{equation}\label{5}
\hat{\rho}_{\phi} = -\, \frac{2}{a^{6}}\,\partial _{\phi }^{2} +
V(\phi)
\end{equation}
corresponds to the energy density of the scalar field in classical
theory (cf. e.g. Ref.~\cite{PR}). The eigenvalue $E$ determines
the components of the energy-momentum tensor $\widetilde T^{0}_{0}
= E/a^{4}$ and $\widetilde T^{\mu}_{\nu} = -\,E/(3\, a^{4})\,
\delta^{\mu}_{\nu}$, where $\mu, \nu = 1,2,3$. We shall consider
the case $E > 0$ and call a source determined by the
energy-momentum tensor $\widetilde T^{\mu}_{\nu}$ a radiation.

In order to find the function $\psi _{E}$ at given $V(\phi)$
Eq.~(\ref{4}) must be supplemented with the boundary condition.
According to Eq.~(\ref{4}) the universe can be both in
quasistationary and continuum states \cite{K}. Quasistationary
states are the most interesting since the universe in such states
can be described by the set of standard cosmological parameters
\cite{KK}. These states are characterized by some complex
parameter $E = E_{n} + i\, \Gamma_{n}$, where $E_{n} > 0$ is a
position, $\Gamma_{n} > 0$ is a width of the $n$-th level, $n =
0,1,2, \ldots$. The wavefunction $\psi _{E}$ of the
quasistationary state as a function of $a$ has a sharp peak and it
is concentrated mainly in the region limited by the barrier $U =
a^{2} - a^{4} V(\phi)$ (see Eq.~(\ref{4})). It can be normalized
\cite{Fo} and used in calculations of expectation values of
operators corresponding to observed parameters within the lifetime
of the universe, when the quasistationary states can be considered
as stationary ones with $E = E_{n}$ (cf. e.g. Ref.~\cite{BZP}).
Such an approximation does not take into account exponentially
small probability of tunneling through the barrier $U$. Below we
shall not go beyond this approximation.

\section{First stage of evolution}
\label{First}

It is convenient to divide the evolution of the universe
conventionally into two stages. Let us assume that at the first
stage the scalar field $\phi$ evolves slowly (in comparison with a
large increase of the average value of the scale factor $\langle a
\rangle$ in the state $\psi_{E}$ normalized as described above)
from some initial state $\phi_{start}$, where $V(\phi_{start})
\sim \rho_{P}$\footnote{It allows us to consider the evolution of
the universe in time in classical sense.}, into a vacuum-like
state with $V(\phi_{vac}) = 0$. During this era from the point of
view of semiclassical description the early universe is filled
with primordial radiation and is charge symmetric. The scalar
field $\phi$ forms a vacuum state with the non-zero energy
density, $V(\phi(t)) \neq 0$, which effectively decreases with
time $t$. At this stage the kinetic term of the operator of the
energy density of the scalar field (\ref{5}) can be neglected
(adiabatic approximation), and it is convenient to represent the
wavefunction of the universe in the $n$-th state, $\psi_{E} =
|E_{n} \rangle$, in the form of expansion in terms of a complete
set of functions $\langle a|n \rangle$ which satisfy the equation
\begin{equation}\label{6}
    \left( -\,\partial _{a}^{2} +  a^{2}  - \epsilon_{n}^{0} \right)
    |n \rangle = 0,
\end{equation}
where $\epsilon_{n}^{0} = 4n + 3$. Then we have
\begin{equation}\label{7}
    |E_{n} \rangle = \sum_{q} |q \rangle \langle q|E_{n} \rangle.
\end{equation}
Taking into account that quasistationary states are realized in
the universe only in the case when $V \ll 1$ \cite{K,KK} and using
the perturbation theory we obtain
\begin{eqnarray}\label{8}
    \langle q |E_{n} \rangle = \delta_{nq}
    - \frac{V}{4}\left[\frac{1}{8}\sqrt{N(N-1)(N-2)(N-3)}\right. \delta_{n-2,q}\\
    + \sqrt{N(N-1)}\left(N- \frac{1}{2}\right) \delta_{n-1,q} -
    \sqrt{(N+1)(N+2)}\left(N+\frac{3}{2}\right) \delta_{n+1,q} \nonumber \\
    -  \left. \frac{1}{8}\sqrt{(N+1)(N+2)(N+3)(N+4)}\,\delta_{n+2,q} \right]-O(V^{2}),
\nonumber
\end{eqnarray}
where $N = 2n + 1$. The eigenvalue $E$ in this approximation is
the following
\begin{equation}\label{9}
    E_{n} = 2N + 1 - \frac{3}{4}\, V\, [2N (N + 1) + 1 ] - O(V^{2}).
\end{equation}
It depends on $\phi$ parametrically. The wavefunction $\langle a|n
\rangle$ describes the geometrical properties of the universe as a
whole. Since in classical theory the gravitational field is
considered as a variation of space-time metric, then this
wavefunction will characterize the quantum properties of the
gravitational field. The states $\langle a|n \rangle$ can be
formally interpreted as those which emerge as a result of motion
of some imaginary particle with the Planck mass $m_{P} =
l_{P}^{-1}$ and zero orbital angular momentum in imaginary field
with the potential energy $U(R) = \frac{1}{2}\,k_{P} R^{2}$, where
$R = l_{P} a$ is a ``radius'' of the curved universe, while
$k_{P}= m_{P}^{3}$ can be called a ``stiffness coefficient of
gravitational field (or space)''. Its numerical value is $k_{P} =
4.79 \times 10^{85}$ GeV $\mbox{cm}^{-2}$. This motion causes the
equidistant spectrum of energy $\mathcal{E}_{n}= m_{P}\left(N +
\frac{1}{2}\right)$, where $m_{P}$ is the energy (mass) of the
elementary quantum excitation of the vibrations of the oscillator
(\ref{6}).

Introducing the operators
\begin{equation}\label{10}
    A^{\dag}= \frac{1}{\sqrt{2}}\,(a - \partial_{a}), \quad
    A = \frac{1}{\sqrt{2}}\,(a + \partial_{a}),
\end{equation}
the state $|n \rangle$ can be represented in the form
\begin{eqnarray}\label{11}
   |n \rangle  =
   \frac{1}{\sqrt{N!}}\,(A^{\dag})^{N}\,|vac \rangle, \quad
   A\,|vac \rangle  = 0, \quad |vac \rangle =
   \left(\frac{4}{\pi}\right)^{1/4}\exp \left\{-
   \frac{a^{2}}{2}\right\}.
\end{eqnarray}
The operators $A^{\dag}$ and $A$ satisfy the ordinary canonical
commutation relations, $[A, A^{\dag}] = 1,\ [A,A] = [A^{\dag},
A^{\dag}] = 0$, and one can interpret them as the operators for
the creation and annihilation of the elementary quantum excitation
with the energy $m_{P}$. The integer $N$ gives the number of these
excitations in the $n$-th state of the universe. The vacuum state
$|vac \rangle$ describes the universe without such excitations.
From Eqs.~(\ref{7}) and (\ref{8}) it follows that the universe can
be characterized by quantum number $n$. In such a description the
gravitational field is considered as a system of the elementary
quantum excitations of the vibrations of the oscillator (\ref{6}).

The interaction between the gravitational field and non-zero
vacuum of the field $\phi$ results in the fact that the
wavefunction (\ref{7}) is a superposition of the states with
different $n$.

When the potential energy density $V(\phi)$ decreases to the value
$V \ll 0.1$ the number of available states of the universe
increases up to $n \gg 1$. By the moment when the scalar field
will roll in the location where $V(\phi_{vac}) = 0$ the universe
\textit{can be found in the state with $n \gg 1$}. This can occur
because the emergence of new quantum levels and the (exponential)
decrease in widthes of old ones result in the appearance of the
competition between the tunneling through the barrier $U$ and
allowed transitions between the states $n \rightarrow n \pm 1,\ n
\pm 2$ (see Eq.~(\ref{8})). A comparison between these processes
demonstrates \cite{K,KK} that the transition $n \rightarrow n + 1$
is more probable than any other allowed transitions or decays. The
vacuum energy in the early universe originally stored by the field
$\phi$ with the potential energy density $V(\phi_{start}) \sim
\rho_{P}$ is a source of transitions with increase in number $n$.

\section{Creation of matter/energy}
\label{Creation}

According to accepted model the scalar field $\phi$ descends to
the state with zero energy density, $V(\phi_{vac}) = 0$. At that
instant the first stage comes to an end and the universe enters
the second stage of its evolution. The main feature of the new era
is a creation of matter/energy which can turn into the ordinary
forms. In the state with $V(\phi_{vac}) = 0$ the field $\phi$
oscillates about the equilibrium vacuum-like state due to quantum
fluctuations. These oscillations can be quantized.

In general case it is convenient to represent the wavefunction
$\psi_{E}$ of Eq.~(\ref{4}) in the form of a superposition of the
states of adiabatic approximation. In the case of the states with
$n \gg 1$ which we shall study the task is simplified. Since the
expansion coefficients of the adiabatic wavefunction (\ref{7})
behave as
\begin{equation}\label{12}
    \langle q|E_{n} \rangle \rightarrow \delta_{nq} \quad
    \mbox{for} \quad n \rightarrow \infty
\end{equation}
up to the terms $\sim O(V^{2})$, then the wavefunction in the
states with $n \gg 1$ in adiabatic approximation coincides with
the function $\langle a|n \rangle$ with above accuracy. And the
desired representation of the wavefunction $\psi_{E}$ has a form
$\psi_{E} = \sum_{n} |n \rangle\, f_{n}$. Multiplying
Eq.~(\ref{4}) by $a^{2}$ on the left and using this expansion, one
obtains the set of equations for the coefficients $f_{n}$ as
functions of $\phi$. In the limit $n \gg 1$ such a set is reduced
to one equation. This equation coincides with the equation which
follows from Eq.~(\ref{4}), if one uses the expansion of the
wavefunction $\psi_{E}$ in terms of a complete set of exact
functions of adiabatic approximation and then passes to limit of
very large numbers $n$ in a final set of equations for the
coefficients of expansion \cite{K}.

We are interested in the states of the field $\phi$ near its
vacuum value $\phi_{vac}$. Therefore it is convenient to pass in
equation for $f_{n}$ from the variable $\phi$ to $x \sim (\phi -
\phi_{vac})$ which characterizes a deviation of $\phi$ from
equilibrium value $\phi_{vac}$. Such an equation has a form
\begin{equation}\label{13}
    \left[\partial_{x}^{2} + z - V(x) \right] f_{n}(x) = 0,
\end{equation}
where $x = \sqrt{m/2}\,(2N)^{3/4}\,(\phi - \phi_{vac})$, $z =
(\sqrt{2N}/m) \left(1 - E/(2N)\right)$, and $V(x) =
(2N)^{3/2}\,V(\phi)/m$, $m$ is some dimensionless parameter. Since
the average value of the scale factor in the state of the universe
with $n \gg 1$ is equal to $\langle a \rangle = \sqrt{N/2}$
\cite{K,KK}, then $V(x)$ is the potential energy of the scalar
field contained in the universe with the volume $\sim \langle a
\rangle^{3}$. The value $x^{2}$ characterizes the deviation
squared of the field $\phi$ in the volume $\sim \langle a
\rangle^{3}$. Therefore Eq.~(\ref{13}) describes the stationary
states which characterize the scalar field $\phi$ in the universe
as a whole.

Choosing the parameter $m^{2} = \left[\partial_{\phi}^{2}
V(\phi)\right]_{\phi_{vac}}$ and expanding $V(\phi)$ in the powers
of $x$ we obtain
\begin{equation}\label{14}
    V(x) = x^{2} + \alpha\,x^{3} + \beta\,x^{4} + \ldots,
\end{equation}
where
\begin{equation}\label{15}
    \alpha = \frac{\sqrt{2}}{3}\, \frac{\lambda}{m^{5/2}}\, \frac{1}{(2N)^{3/4}},
    \quad
    \beta = \frac{1}{6}\, \frac{\nu}{m^{3}}\, \frac{1}{(2N)^{3/2}},
\end{equation}
and $\lambda =
\left[\partial_{\phi}^{3}V(\phi)\right]_{\phi_{vac}}$, $\nu =
\left[\partial_{\phi}^{4}V(\phi)\right]_{\phi_{vac}}$. The
potential energy density $V(\phi)$ near the point $\phi_{vac}$
will be considered to be smooth sufficiently so that $m^{2} >
\lambda > \nu$.

Since $N \sim n \gg 1$, then $|\alpha| \ll 1$, $|\beta| \ll 1$,
and Eq.~(\ref{13}) can be solved using the perturbation theory for
stationary problems with a discrete spectrum. We take for the
state of the unperturbed problem the state of the harmonic
oscillator with the equation of motion (\ref{13}), where $V(x) =
x^{2}$. In the occupation number representation one can write for
the unperturbed states
\begin{eqnarray}\label{16}
   f_{ns}^{0} = \frac{1}{\sqrt{s!}}\,(B_{n}^{\dag})^{s} f_{n0}^{0},
   \quad B_{n}\,f_{n0}^{0} = 0, \quad f_{n0}^{0} (x) =
   \left(\frac{1}{\pi}\right)^{1/4}\exp \left\{-
   \frac{x^{2}}{2}\right\}
\end{eqnarray}
with $z^{0} = 2s + 1$, where $s = 0, 1, 2 \ldots$, and it is
denoted
\begin{equation}\label{17}
   B_{n}^{\dag}= \frac{1}{\sqrt{2}}\,(x - \partial_{x}),
   \qquad
   B_{n} = \frac{1}{\sqrt{2}}\,(x + \partial_{x}).
\end{equation}
Solving Eq.~(\ref{13}) with the potential (\ref{14}) we find
\begin{equation}\label{18}
    z = 2 s + 1 + \Delta z,
\end{equation}
where
\begin{eqnarray}\label{19}
    \Delta z & = & \frac{3}{2} \beta \left(s^{2} + s + \frac{1}{2}
    \right) - \frac{15}{8} \alpha^{2} \left(s^{2} + s + \frac{11}{30}
    \right) \nonumber \\
    & - & \frac{\beta^{2}}{16} \left(34 s^{3} + 51 s^{2} + 59 s +
    21 \right).
\end{eqnarray}
The spectrum of the energy states of the field $\phi$ has a form
\begin{equation}\label{20}
    M_{s} = m \left(s + \frac{1}{2}\right) + \Delta M,
\end{equation}
where $\Delta M = m \Delta z/2$. The operators $B_{n}^{\dag}$ and
$B_{n}$ satisfy the ordinary canonical commutation relations,
$[B_{n}, B_{n}^{\dag}] = 1,\ [B_{n},B_{n}] = [B_{n}^{\dag},
B_{n}^{\dag}] = 0$, and can be interpreted as the creation
($B_{n}^{\dag}$) and annihilation ($B_{n}$) operators which
increase and decrease the number of elementary quantum excitations
of the vibrations of the scalar field by unity in the universe in
the $n$-th state, respectively. Therefore it is natural to
interpret the value $M = m (s + 1/2)$ as a quantity of
matter/energy being the sum of elementary quantum excitations of
the vibrations of the field $\phi$ with the energy (mass) $m$,
while $s$ counts the number of these excitations. It can be
considered as additional quantum number. The summand $\Delta M$ in
Eq.~(\ref{20}) takes into account a self-action of the elementary
quantum excitations of the vibrations of the field $\phi$.

Using Eq.~(\ref{18}) and the relation between $z$ and $E$ we find
the condition of quantization
\begin{equation}\label{21}
    E = 2 N - \sqrt{2 N} M_{s}.
\end{equation}
This condition can be rewritten in terms of classical cosmology in
the form of the relation between the parameters of the universe
\begin{equation}\label{22}
    \langle a \rangle = M_{s} + \frac{E}{4 \langle a \rangle},
\end{equation}
where $\langle a \rangle$ is the scale factor, $M_{s}$ is the
energy of matter (in the form of a system of elementary quantum
excitations of the vibrations of the scalar field), while $E/(4
\langle a \rangle)$ is the energy of radiation.

Let us estimate $\Delta M/M$. Substituting Eq.~(\ref{15}) into
(\ref{19}), leaving the main terms only, and  taking in accordance
with Eq.~(\ref{22}) that $\langle a \rangle \sim m s$ we obtain
\begin{equation}\label{24}
    \frac{\Delta M}{M} \sim \left[0.02\,\frac{\nu}{m^{2}} -
    0.03\,\left(\frac{\lambda}{m^{2}}\right)^{2}\right]\frac{1}{(m^{2}s)^{2}}
    - 4.6 \times 10^{-4} \left(\frac{\nu}{m^{2}}\right)^{2}
    \frac{1}{(m^{2}s)^{4}}.
\end{equation}
From here it follows that
\begin{equation}\label{25}
    \frac{\Delta M}{M} \ll 1 \quad \mbox{at} \quad s > m^{-2}.
\end{equation}

Let us estimate the value $s$ at which condition of smallness of
$\Delta M/M$ is satisfied. In the limit of maximum possible mass
$m \sim 1\, (\sim 10^{19}\, \mbox{GeV})$ we find $s > 1$. Such a
universe has the parameters: $M > 1\, (\sim 10^{19}\,
\mbox{GeV})$, $\langle a \rangle > 1\, (\sim 10^{-33}\,
\mbox{cm})$ and age $t > 1\, (\sim 10^{-44}\, \mbox{s})$. Taking
$s \sim 10^{80}$ (equivalent number of baryons in the present-day
universe) we obtain the limitation on mass from below, $m >
10^{-40}\, (\sim 10^{-21}\, \mbox{GeV})$. Assuming that $m \sim
10^{-18}\, (\sim 10\, \mbox{GeV})$ we find the following
restriction on the number of elementary quantum excitations of the
vibrations of the field $\phi$: $s > 10^{36}$. The quantity of
matter/energy in such a universe is $M > 10^{18}\, (\sim 10^{37}\,
\mbox{GeV})$, radius of curvature $\langle a \rangle > 10^{18}\,
(\sim 10^{-15}\,\mbox{cm})$, and age $t > 10^{-26}$ s. Below (in
Sec.~5) we shall consider the model of creation of ordinary matter
which leads to $m \gtrsim 10$ GeV.

At the end of this section let us calculate the mean energy
density in the universe in the states with large quantum numbers.
From Eq.~(\ref{4}) it follows that the operator
\begin{equation}\label{26}
    \hat{\rho}_{tot} = \hat{\rho}_{\phi} + \frac{E}{a^{4}}
\end{equation}
corresponds to the total energy density in classical theory. Then
the average value $\langle \hat{\rho}_{tot} \rangle$ in the state
of the universe with $n \gg 1$, $s \gg 1$ is
\begin{equation}\label{27}
    \langle \hat{\rho}_{tot} \rangle = \gamma \frac{M}{\langle a
    \rangle^{3}} + \frac{E}{\langle a \rangle^{4}},
\end{equation}
where $\gamma = 193/12$ and it is assumed that the average value
$\langle a \rangle$ determines the scale factor of the universe in
semiclassical description (details can be found in
Refs.~\cite{K,KK,KK4}). In this approximation the universe is
described by the Einstein-Friedmann equations in terms of average
values which follow from Eq.~(\ref{4}) and from Heisenberg-type
equation. The latter determines a change in time of the average
values of the physical quantities \cite{KK4}. In the matter
dominated universe $M \gg E/(4 \langle a \rangle)$ and the mean
energy density (\ref{27}) leads to the dimensionless density
$\Omega_{tot} \equiv \langle \hat{\rho}_{tot} \rangle/H^{2} =
1.066$, where $\langle \hat{\rho}_{tot} \rangle$ is measured in
units of $\rho_{P}$ and the Hubble constant $H$ in inverse Planck
time $t_{P}^{-1} = l_{P}^{-1}$. It means that the universe in
highly excited states is very close to being spatially flat. It
agrees with existing astrophysical data for the present-day
universe (see Sec.~1). Moreover a very slight systematical excess
of $\Omega_{tot}$ over unity is observed \cite{Sp}.

\section{Dark matter and baryonic matter production}
\label{Production}

The elementary quantum excitations of the vibrations of the field
$\phi$ are subject to action of gravity. Due to this fact they can
decay into the ordinary particles (e.g. baryons and leptons) that
have to be present in the universe in small amount because of the
weakness of the gravitational interaction.

As it is mentioned in Sec.~1 the total energy density can be
represented as a sum of three terms
\begin{equation}\label{28}
    \Omega_{tot} = \Omega_{B} + \Omega_{DM} + \Omega_{X}.
\end{equation}
Therefore in simple (naive) model it is natural to assume that the
elementary quantum excitation (let us denote it by $\phi$ as well)
of the vibrations of the scalar field decays according to a scheme
\begin{eqnarray}\label{29}
   \phi \longrightarrow \phi' + \nu\, + \!\!\!\! &n& \nonumber \\[-0.1cm]
   &\rvert & \nonumber \\[-0.3cm]
   & &\!\!\!\!\!\! \longrightarrow p + e^{-} + \overline{\nu}
\end{eqnarray}

\noindent with stable particles in the final state. Here $\phi'$
is the quantum of the residual excitations of the vibrations of
the scalar field. A system of such quanta reveals itself in the
universe in the form of non-baryonic dark matter. Neutrino $\nu$
takes away the spin, and the dark matter particle is a boson. At
the first stage of decay chain (\ref{29}) the baryon number is not
conserved but the spin and the energy are conserved. A system of
the elementary quantum excitations of the vibrations of the scalar
field can be interpreted as dark energy (see below). Such
elementary quantum excitations we shall call dark energy quanta
for briefness. Assuming that dark energy quanta decay
independently, as well as neutrons $n$, one can write a set of
equations for decay chain (\ref{29}) as follows
\begin{eqnarray}\label{30}
  \frac{ds(t)}{dt}\!\!\! &=&\!\!\! - \Gamma_{\phi}(t)\,s(t),\quad
  \frac{ds_{n}(t)}{dt} = - \Gamma_{n}(t)\,s_{n}(t) + \Gamma_{\phi}(t)\,s(t),\nonumber \\
  \frac{ds_{p}(t)}{dt}\!\!\! &=&\!\!\! - \Gamma_{p}(t)\,s_{p}(t) +
  \Gamma_{n}(t)\,s_{n}(t),
\end{eqnarray}
where $s(t)$ is the number of dark energy quanta as a function of
time $t$, $s_{n}(t)$ and $s_{p}(t)$ are the numbers of neutrons
and protons at some instant of time $t$, $\Gamma_{\phi}(t)$ and
$\Gamma_{n}(t)$ are the decay rates of dark energy quantum and
neutron respectively, while the decay rate of proton will be
supposed to be zero below, $\Gamma_{p}(t) = 0$. The initial
conditions have the form
\begin{equation}\label{31}
    s(t') = s, \quad s_{n}(t') = 0, \quad s_{p}(t') = 0,
\end{equation}
where $s$ is the number of dark energy quanta at some arbitrary
chosen initial instant of time $t'$. The first equation of the set
(\ref{30}) describes the exponential law of decrease of the number
of dark energy quanta with time,
\begin{equation}\label{32}
    s(t) = s\,\mbox{e}^{-\overline{\Gamma}_{\phi}\,\Delta t},
\end{equation}
where
\begin{equation}\label{33}
    \overline{\Gamma}_{\phi} = \frac{1}{\Delta t} \int_{t'}^{t}\!
    dt\, \Gamma_{\phi}(t)
\end{equation}
is the mean decay rate on the interval $\Delta t = t - t'$.
Substituting (\ref{32}) into the second equation of the set
(\ref{30}) and taking into account (\ref{31}) we obtain the law of
production of neutrons
\begin{equation}\label{34}
    s_{n}(t) = s \int_{t'}^{t}\! d\tau\,\Gamma_{\phi}(\tau)\,\exp
    \left\{-\int_{t'}^{\tau}\! dt''\,\Gamma_{\phi}(t'') -
    \int_{\tau}^{t}\! dt''\,\Gamma_{n}(t'') \right\}.
\end{equation}
The number of protons will be
\begin{equation}\label{35}
    s_{p}(t) = s \int_{t'}^{t}\! d\tau\,\Gamma_{n}(\tau)\,
     \int_{t'}^{\tau}\! d\tau'\,\Gamma_{\phi}(\tau')\,\exp
    \left\{-\int_{t'}^{\tau'}\! dt''\,\Gamma_{\phi}(t'') -
    \int_{\tau'}^{\tau}\! dt''\,\Gamma_{n}(t'') \right\}.
\end{equation}

The decay rates $\Gamma_{n}(t)$ and $\Gamma_{\phi}(t)$ in
Eqs.~(\ref{34}) and (\ref{35}) are unknown. Let us assume that in
the decay (\ref{29}) free particles are produced and the energy is
released. Then we can write energy balance equation for the decay
of dark energy quantum with the energy $m$
\begin{equation}\label{36}
    m = m_{\phi'} + m_{n} + m_{\nu} + Q,
\end{equation}
where $m_{\phi'}$, $m_{n}$, and $m_{\nu}$ are the masses of dark
matter particle, neutron and neutrino, respectively, and
\begin{equation}\label{37}
    Q = \sum_{i}\left(\sqrt{m_{i}^{2} + p_{i}^{2}} - m_{i}
    \right),\quad i = \{\phi',n,\nu\},
\end{equation}
is the energy released in the decay, $p_{i}$ is the momentum of
the particle $i$. The summand $m_{\nu}$ in Eq.~(\ref{36}) we shall
include in $m_{\phi'}$ considering neutrinos as a constituent part
of dark matter.

At high temperatures $T \sim Q$ (small ages $t$) the rate
$\Gamma_{n}$ is proportional to fifth power of temperature,
$\Gamma_{n} \sim T^{5}$. When the temperature decreases (during
the expansion of the universe) the rate $\Gamma_{n}$ decreases as
well and at low temperatures ($t \gg 1$ s) it tends to the mean
decay rate of free neutron, $\overline{\Gamma}_{n} = 1.12 \times
10^{-3}\ \mbox{s}^{-1}$. Therefore for estimation it is enough to
take as $\Gamma_{n}(t)$ its smallest value
$\overline{\Gamma}_{n}$. Moreover we shall assume that the mean
decay rate of dark energy quantum (\ref{33}) depends very weakly
on averaging interval. Then in indicated approximation from
Eqs.~(\ref{34}) and (\ref{35}) we obtain the simple expressions
\begin{equation}\label{38}
  \frac{s_{n}(t)}{s} = \frac{\overline{\Gamma}_{\phi}}{\overline{\Gamma}_{n}
  - \overline{\Gamma}_{\phi}} \left(\mbox{e}^{- \overline{\Gamma}_{\phi} \Delta t}
  - \mbox{e}^{- \overline{\Gamma}_{n} \Delta t}\right),
\end{equation}
\begin{equation}\label{39}
  \frac{s_{p}(t)}{s} = 1 + \frac{1}{\overline{\Gamma}_{n} -
  \overline{\Gamma}_{\phi}} \left( \overline{\Gamma}_{\phi}
   \mbox{e}^{- \overline{\Gamma}_{n} \Delta t} -
   \overline{\Gamma}_{n}\mbox{e}^{- \overline{\Gamma}_{\phi}
   \Delta t}\right).
\end{equation}
It is easy to make sure that the law of conservation of number of
particles, $s = s(t) + s_{n}(t) + s_{p}(t)$, holds at every
instant of time $t$. Since we are interested in matter/energy
density in the present-day universe, then for numerical
estimations we choose $\Delta t$ equal to the age of the universe,
$\Delta t = 14$ Gyr \cite{PR,Kr}. In this case
$\overline{\Gamma}_{n} \Delta t = 5 \times 10^{14} \gg 1$. We
suppose that the decay of dark energy quantum is caused
\textit{mainly} by the action of gravitational forces, so that
$\overline{\Gamma}_{\phi} \ll \overline{\Gamma}_{n}$. Then
Eq.~(\ref{39}) is simplified
\begin{equation}\label{40}
    s_{p}(t) = \overline{s} = s\, \left( 1 - \mbox{e}^{- \overline{\Gamma}_{\phi}
   \Delta t}\right),
\end{equation}
where $\overline{s}$ is an average number of dark energy quanta
which decay during the time interval $\Delta t$. Since dark matter
particles $\phi'$ are assumed to be stable (with lifetime greater
than $\Delta t$) their number is equal to $\overline{s}$ as well.
The mean decay rate $\overline{\Gamma}_{\phi}$ is unknown and must
be calculated on the basis of vertex modelling of complex decay
(\ref{29}) or extracted from astrophysical data.

According to Eq.~(\ref{27}) in the matter dominated universe the
total energy density $\Omega_{tot}$ for sufficiently large number
$s$ of dark energy quanta equals
\begin{equation}\label{41}
    \Omega_{tot} = \gamma \, \frac{m s}{\langle a \rangle^{3}
    H^{2}}.
\end{equation}
The densities of (both optically bright and dark) baryons and dark
matter are equal to
\begin{equation}\label{42}
    \Omega_{B} = \gamma \, \frac{m_{p}\, \overline{s}}{\langle a \rangle^{3}
    H^{2}}, \quad
    \Omega_{DM} = \gamma \, \frac{m_{\phi'}\, \overline{s}}{\langle a \rangle^{3}
    H^{2}},
\end{equation}
where $m_{p} = 0.938$ GeV is a proton mass. Since the density of
visible baryons is equal to a ratio of total mass to volume, then
in accepted units
\begin{equation}\label{43}
    \Omega_{stars} = \frac{m_{p}\, \overline{s}}{\langle a \rangle^{3}
    H^{2}}.
\end{equation}
Then from Eqs.~(\ref{42}) and (\ref{43}) it follows that the
coefficient $\gamma$ determines a ratio between the densities
\begin{equation}\label{44}
    \frac{\Omega_{B}}{\Omega_{stars}} \simeq 16.08.
\end{equation}
On the order of magnitude this value agrees with the observational
data (see Sec.~1).

Taking into account Eqs.~(\ref{28}), (\ref{41}), (\ref{42}) and
(\ref{43}) we obtain the following expressions for the energy
density components
\begin{equation}\label{45}
    \frac{\Omega_{B}}{\Omega_{tot}} =
    \frac{m_{p}}{m}\,\frac{\overline{s}}{s},\quad
    \frac{\Omega_{DM}}{\Omega_{tot}} = \left(1 -
    \frac{Q + m_{n}}{m}\right)\,\frac{\overline{s}}{s},
\end{equation}
and
\begin{equation}\label{46}
    \frac{\Omega_{M}}{\Omega_{tot}} = \left(1 -
    \frac{Q + \Delta m}{m}\right)\,\frac{\overline{s}}{s},\quad
    \frac{\Omega_{X}}{\Omega_{tot}} = 1 -
    \frac{\Omega_{M}}{\Omega_{tot}},
\end{equation}
where $\Delta m = m_{n} - m_{p} = 1.293$ MeV. All components are
expressed in terms of three unknown parameters:
$\overline{\Gamma}_{\phi}$, $m$ and $Q$.

Let us introduce the dimensionless gravitational coupling constant
$g = G m^{2}$ for a particle with mass $m$. Then using
Eq.~(\ref{40}) the baryonic component can be rewritten as
\begin{equation}\label{47}
    \frac{\Omega_{B}}{\Omega_{tot}} = \sqrt{\frac{g_{p}}{g}}\,
    \left( 1 - \mbox{e}^{- \overline{\Gamma}_{\phi} \Delta t}\right),
\end{equation}
where $g_{p} = 0.590 \times 10^{-38}$ is the gravitational
coupling constant for a proton. In order to find the possible form
of $\overline{\Gamma}_{\phi}$ as a function of $g$ let us use an
analogy and take as an example the rate of decay of some particle
$B$ into a pair of leptons (see e.g. Ref.~\cite{LP})
\begin{equation}\label{48}
    \Gamma(B \rightarrow e^{+} e^{-}) = \frac{16
    \pi}{3}\,\alpha^{2}\,\frac{|\psi(0)|^{2}}{\mu^{2}},
\end{equation}
where $\alpha$ is the dimensionless fine-structure constant, $\mu$
is the mass of a particle $B$, $\psi(0)$ is its wavefunction at
the origin. The factor $|\psi(0)|^{2}$ is a particle number
density and the value $L \sim |\psi(0)|^{-2/3}$ gives a linear
dimension of an area from which the pair $e^{+} e^{-}$ is emitted.
On the order of magnitude $L$ characterizes a size of a particle
$B$. Making substitutions $\alpha \rightarrow g$, $\mu \rightarrow
m$ we obtain the expression for $\overline{\Gamma}_{\phi}$
\begin{equation}\label{49}
    \overline{\Gamma}_{\phi} = 8 \pi^{2} g\, |\psi(0)|^{2},
\end{equation}
where $\psi(0)$ is the wavefunction of the dark energy quantum at
the origin.

According to Eqs.~(\ref{47}) and (\ref{49}) for fixed $\Delta t$
the density $\Omega_{B}$ is the function of $g$. It vanishes at $g
= 0$ and tends to zero as $g^{-1/2}$ at $g \rightarrow \infty$. It
has one maximum. Let us fix the coupling constant $g$ by maximum
value of $\Omega_{B}(g)$. Then we obtain
\begin{equation}\label{50}
    \overline{\Gamma}_{\phi}\, \Delta t = 1.256.
\end{equation}
For $\Delta t = 14$ Gyr it gives
\begin{equation}\label{51}
    \overline{\Gamma}_{\phi} = 2.840 \times 10^{-18}\
    \mbox{s}^{-1}.
\end{equation}
This rate satisfies inequality $\overline{\Gamma}_{\phi} \ll
\overline{\Gamma}_{n}$, and
\begin{equation}\label{52}
    \overline{\Gamma}_{\phi} > H_{0},
\end{equation}
where $H_{0} = 71\ \mbox{km s}^{-1}\mbox{Mpc}^{-1}$ is the
present-day value of the Hubble expansion rate \cite{Kr,Gr}. The
latter condition means that on average at least one interaction
has occurred over the lifetime of the universe.

Substituting Eq.~(\ref{50}) into (\ref{40}) we find
\begin{equation}\label{53}
    \frac{\overline{s}}{s} = 0.715,
\end{equation}
i.e. about $70$\% of all elementary excitations of the vibrations
of the primordial scalar field had to decay during the elapsed
time $\Delta t = 14$ Gyr.

If one knows $\overline{\Gamma}_{\phi}$ and $\psi(0)$, then using
Eq.~(\ref{49}) it is possible, in principle, to restore the
coupling constant $g$. But the wavefunction of dark energy
quantum, as well as an equation it must satisfy, is unknown.
Therefore let us consider an inverse problem. Namely, using the
observed value of $\Omega_{B}$ we shall restore $g$ and then
obtain all masses and density components. For definiteness we
choose $\Omega_{B}/\Omega_{tot} = 0.04$. Then from Eq.~(\ref{44})
we find the density of visible baryons
\begin{equation}\label{54}
    \frac{\Omega_{stars}}{\Omega_{tot}} = 0.0025.
\end{equation}
For a flat universe this value is in good agreement with
observations (see Sec.~1). Then using Eqs.~(\ref{45}) and
(\ref{53}) we find
\begin{equation}\label{55}
    g \simeq 320\,g_{p}, \qquad m \simeq 16.8\, \mbox{GeV}.
\end{equation}
The restriction on the energy $Q$ follows from the condition
$m_{\phi'} \geq 0$
\begin{equation}\label{56}
    0 \ \mbox{GeV} \leq Q \leq 15.8 \ \mbox{GeV}
\end{equation}
with central value $\overline{Q} = 7.9$ GeV. It, in turn, leads to
following restrictions on the mass of dark matter particle and on
the density components
\begin{equation}\label{57}
    0 \ \mbox{GeV} < m_{\phi'} < 15.8 \ \mbox{GeV}
\end{equation}
with central value $\overline{m}_{\phi'} = 7.9$ GeV,
\begin{equation}\label{58}
    0 < \frac{\Omega_{DM}}{\Omega_{tot}} < 0.67
\end{equation}
with central value $\overline{\Omega}_{DM}/\Omega_{tot} \simeq
0.34$,
\begin{equation}\label{59}
    0.04 < \frac{\Omega_{M}}{\Omega_{tot}} < 0.71
\end{equation}
with central value $\overline{\Omega}_{M}/\Omega_{tot} \simeq
0.38$,
\begin{equation}\label{60}
    0.96 > \frac{\Omega_{X}}{\Omega_{tot}} > 0.29
\end{equation}
with central value $\overline{\Omega}_{X}/\Omega_{tot} \simeq
0.62$. Here the left-hand sides of the inequalities correspond to
$Q = 15.8$ GeV, while the right-hand sides to $Q = 0$ GeV. From
inequality (\ref{60}), in particular, it follows that if
practically all energy of the elementary quantum excitation of the
vibrations of the scalar field transforms into the energy $Q$,
then $\Omega_{X} \sim \Omega_{tot}$.

The central values of the density components
$\overline{\Omega}_{DM}$ and $\overline{\Omega}_{M}$ mentioned
above are undoubtedly overestimated, since they take into account
the upper limits for these components corresponding to unlikely
value $Q = 0$ GeV. In naive model (\ref{29}) under consideration
it make sense to speak only about the order of magnitude of the
mean energy $Q$, which according to Eq.~(\ref{56}) is equal to $Q
\simeq 10$ GeV. The following parameters correspond to such an
energy
\begin{equation}\label{61}
    \frac{\Omega_{DM}}{\Omega_{tot}} \simeq 0.25, \quad
    \frac{\Omega_{M}}{\Omega_{tot}} \simeq 0.29, \quad
    \frac{\Omega_{X}}{\Omega_{tot}} \simeq 0.71,
\end{equation}
and $m_{\phi'} \simeq 5.8$ GeV. In Figs.~1 and 2 the theoretical
values of densities $\Omega_{M}/\Omega_{tot}$ (\ref{59}) and
$\Omega_{X}/\Omega_{tot}$ (\ref{60}) in comparison with
observational data summarized in Ref.~\cite{Sp} are shown. There
is a good agreement between combined observational data (Fig.~2)
and the theoretical prediction (point D corresponding to the case
$Q \simeq 10$ GeV).

\begin{figure}[ht]
\begin{center}
\includegraphics*[scale=1.4]{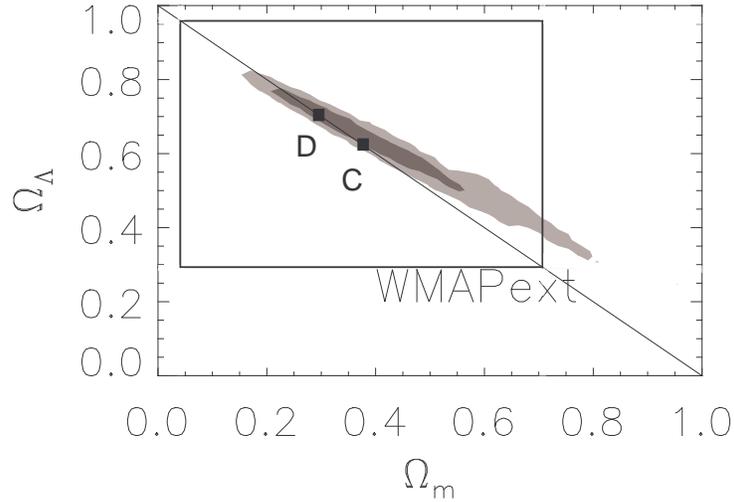}
\end{center}
\caption{The plane $\Omega_{\Lambda} \equiv
\Omega_{X}/\Omega_{tot}$ vs. $\Omega_{m} \equiv
\Omega_{M}/\Omega_{tot}$. Constraints on the density components
determined using WMAP + other CMB experiments (from
Ref.~\cite{Sp}). The acceptable values of $\Omega_{\Lambda}$ and
$\Omega_{m}$ (in accordance with inequalities (\ref{59}) and
(\ref{60})) lie on the diagonal of rectangle. The central value of
the region is shown as a solid box C. The point D corresponds to
the case $Q \simeq 10$ GeV.} \label{fig:1}
\end{figure}

\begin{figure}[ht]
\begin{center}
\includegraphics*[scale=1.4]{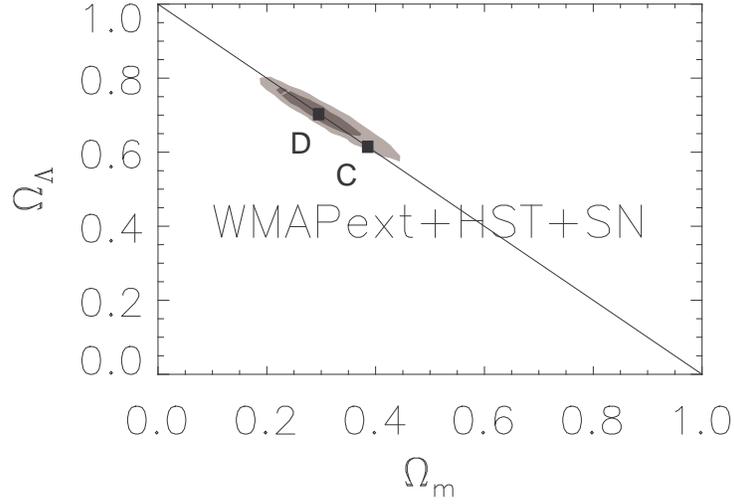}
\end{center}
\caption{Constraints on the density of matter $\Omega_{m}$ and
dark energy $\Omega_{\Lambda}$ determined using WMAPext + HST key
project data + supernova data (from Ref.~\cite{Sp}). The rest as
in Fig.~1.} \label{fig:2}
\end{figure}

Substituting the decay rate (\ref{51}) (multiplied  on the Planck
time $t_{P}$) and the coupling constant $g$ from Eq.~(\ref{55})
into Eq.~(\ref{49}) we obtain the following value for the dark
energy quantum number density
\begin{equation}\label{62}
    |\psi(0)|^{2} \simeq \left(10^{-24}\,\mbox{cm} \right)^{-3}.
\end{equation}
This estimation demonstrates that the decay of the elementary
quantum excitation of the vibrations of the scalar field according
to a scheme (\ref{29}) occurs in an area with linear dimension $L
\sim 10^{-24}$ cm corresponding to the energy $\sim 2 \times
10^{10}$ GeV.

\section{Conclusion}
\label{Conclusion}

According to the model under consideration the universe in the
states with large quantum numbers can be described by the
superposition of quantum states of two oscillators. One oscillator
describes gravitational component as a system of massive
elementary quantum excitations related to the vibrations of
geometry. Another oscillator describes elementary quantum
excitations of the vibrations of primordial matter represented by
the uniform scalar field. The latter excitations are spatially
homogeneous and they form nonluminous (dark) energy. Mainly under
the action of gravitational forces elementary quantum excitations
of the vibrations of the scalar field (dark energy quanta) decay
and produce non-baryonic dark matter, optically bright and dark
baryons, and leptons. Approximately 2/3 of the total energy of all
dark energy quanta has to transform into masses and energies of
observed particles and dark matter up to now. The energy $Q$
released in decay of one dark energy quantum is a free parameter
of the model. The possible values of $Q$ determine the limits of
variations of the densities of dark matter $\Omega_{DM}$ and dark
energy $\Omega_{X}$. These densities turn out to be the values of
the same order of magnitude. The numerical estimations for flat
universe lead to realistic (observed) values of both the matter
density $\Omega_{M} \simeq 0.29$ (with the contributions from dark
matter, $\Omega_{DM} \simeq 0.25$, and optically bright baryons,
$\Omega_{stars} \simeq 0.0025$) and the dark energy density
$\Omega_{X} \simeq 0.71$ if one takes that the mean energy $\sim
10$ GeV is released in separate event of decay of one dark energy
quantum and fixes baryonic component $\Omega_{B} = 0.04$ according
to observational data. The energy (mass) of dark energy quantum is
equal to $\sim 17$ GeV, while the energy $\gtrsim 2 \times
10^{10}$ GeV is needed in order to detect it. Dark matter particle
has the mass $\sim 6$ GeV and such a dark matter has to be
classified as cold.

Let us discuss some consequences of the model under consideration.

\subsection{The parameters of the early universe}
\label{Parameters}

In the radiation dominated universe the mean energy density equals
$\langle \hat{\rho}_{tot} \rangle = \rho_{rad}$, where
\begin{equation}\label{63}
    \rho_{rad} = \frac{2 \pi^{4}}{15}\,N(T)\,T^{4}
\end{equation}
is the energy density of radiation, $N(T)$ counts the total number
of effectively massless degrees of freedom \cite{KK,ZN,Ol2}. Using
the definition $\Omega_{tot} = \langle \hat{\rho}_{tot} \rangle /
H^{2}$ we find the relation between the Hubble constant $H$ and
the temperature $T$
\begin{equation}\label{64}
    H = 2\,\pi^{2} \left(\frac{N(T)}{30\,\Omega_{tot}}\right)^{1/2}
    T^{2}.
\end{equation}
The quantum model predicts the following relation between the age
of the universe $t$ and the Hubble constant $H$: $H t = 1$
\cite{KK4}. This equation explains the observed value of the
dimensionless age parameter $H_{0} t_{0}$ for the present-day
universe, $0.72 \lesssim H_{0} t_{0} \lesssim 1.17$ \cite{PR},
$H_{0} t_{0} = 0.96 \pm 0.04$ \cite{To} and $H_{0} t_{0} \simeq
0.93$ \cite{Kr}. Let us note that standard classical cosmology
\cite{Ol2,We} leads to the relation $H t \rightarrow \frac{1}{2}$
as $t \rightarrow 0$ which gives on the order of magnitude the
correct value of the age of the early universe. The
temperature-time relationship at early times can be written as
\begin{equation}\label{65}
    T = \left(\frac{\Omega_{tot}}{13 N(T)}\right)^{1/4}
    \frac{1}{\sqrt{t}}.
\end{equation}

We shall estimate the temperature of dark matter particles,
baryons,  and leptons which were produced in the process
(\ref{29}). Let us assume that the mean energy per particle in hot
plasma is about $3\, T$ \cite{ZN}. Then for the decay energy $Q
\simeq 10$ GeV the temperature of matter consisting of particles
of decay (\ref{29}) will be equal to $T \simeq 0.67$ GeV ($\sim
0.8 \times 10^{13}$ K). The effective number of relativistic
degrees of freedom for this temperature is equal to $N(T) \simeq
70$ according to the Standard Model of particle physics
\cite{SWO,KT}. Then using Eq.~(\ref{65}) we obtain that the age of
the universe in thermal equilibrium with the temperature $T \simeq
0.67$ GeV is equal to $t \simeq 10^{-6}$ s. The curvature radius
may reach the values $\langle a \rangle \sim 10^{38}\, (\sim
10^{5}\,\mbox{cm})$ for the expansion law $\langle a \rangle \sim
t$ \cite{KK4} or $\langle a \rangle \sim 10^{19}\, (\sim
10^{-14}\,\mbox{cm})$ for $\langle a \rangle \sim \sqrt{t}$
\cite{We}.

It easy to see that the quantum numbers $n \sim \langle a
\rangle^{2}$ and $s > m^{-2}$ satisfy the validity condition of
the model, $n \gg 1$ and $s \gg 1$ for $m \sim 20\, \mbox{GeV}\
(\sim 10^{-18})$.

\subsection{The Jeans mass}
\label{Jeans}

We shall consider the influence of dark matter which consists of
particles with masses $m_{\phi'} \simeq 6$ GeV on the formation of
large-scale structure of the universe. According to standard
theory of large-scale structure formation (see e.g.
Refs.~\cite{ZN,We,Pe}) the gravitational instability boundary is
determined by the Jeans mass $M_{J}$. It is the mass of matter for
which pressure and gravitational attraction compensate each other.

Let us calculate the Jeans mass for dark matter which we shall
consider as a gas of particles with masses $m_{\phi'}$. We shall
assume that in the early radiation dominated universe a
temperature of dark matter was distributed almost uniformly and
was equal to radiation temperature $T$. Since the number density
of dark matter particles $n_{\bar{s}}$ is equal to the number
density of baryons $n_{B}$ according to Eq.~(\ref{40}),
$n_{\bar{s}} = n_{B}$, then the dark matter energy density can be
written as
\begin{equation}\label{66}
    \rho_{\phi'} = 8\, \zeta(3)\, \eta\, m_{\phi'}\, T^{3},
\end{equation}
where $\zeta(3) = 1.2021$, the mass $m_{\phi'}$ and temperature
$T$ are measured in units of $m_{P}$, while $\rho_{\phi'}$ in
units of $\rho_{P}$. Here $\eta = n_{B}/n_{\gamma}$, where
$n_{\gamma}$ is the number density of photons.

The dimensionless Jeans wavelength in the case under consideration
is
\begin{equation}\label{67}
    \lambda_{J} = \frac{\pi}{\sqrt{3\, \zeta(3)\, \eta}}\,
    \frac{1}{m_{\phi'}\, T}.
\end{equation}
Substituting Eqs.~(\ref{66}) and (\ref{67}) into the definition of
the Jeans mass
\begin{equation}\label{68}
    M_{J} = \frac{1}{3 \pi}\, \lambda_{J}^{3}\, \rho_{\phi'},
\end{equation}
where $M_{J}$ is measured in units of $m_{P}$ we obtain
\begin{equation}\label{69}
    M_{J} = \frac{4.62}{\sqrt{\eta}\, m_{\phi'}^{2}}.
\end{equation}
Passing to the ordinary physical units we have
\begin{equation}\label{70}
    M_{J} = \frac{0.77 \times 10^{7}}{\sqrt{\eta_{10}}}\,
    \frac{M_{\odot}}{m_{\phi'}^{2}(GeV)},
\end{equation}
where we introduce the standard notation $\eta_{10}\, \equiv
10^{10} \eta$, $M_{\odot} = 1.12 \times 10^{57}$ GeV is a solar
mass, and $m_{\phi'}$ is measured in units of GeV. The
light-element abundances show that the parameter $\eta_{10}$ lies
in the range between 1 and 10 \cite{Ol,FS}. Therefore from
Eq.~(\ref{70}) for $m_{\phi'} \simeq 6$ GeV we find
\begin{equation}\label{71}
    M_{J} \sim 10^{5} M_{\odot}.
\end{equation}
This estimation demonstrates that if dark matter with above
mentioned properties exists in the universe, then the growth of
non-homogeneities starts from the mass which is $10$ times smaller
then the mass of globular cluster and $10^{6}$ times smaller then
the mass of typical galaxy. The mass $M_{J}$ (\ref{70}) does not
depend on the value of the temperature $T$ in the model under
consideration with uniform distribution of $T$. The mass $M_{J}$
is determined by a ratio of the number density of baryons to
photons $\eta$ and by the mass of dark matter particles
$m_{\phi'}$. The estimation (\ref{71}) holds before hydrogen
recombination. After recombination the evolution of structures
with masses greater than $M_{J}$ (\ref{71}) can be considered
disregarding pressure \cite{ZN}.

The Jeans mass $M_{J}$ (\ref{71}) demonstrates that structures
like globular clusters must form first aggregating to form larger
structures (galaxies and so on) later. It is known that the
cosmology with cold dark matter particles (with mass $> 1$ MeV)
reproduces the observed large-scale structure of the universe much
better than the cosmology with hot dark matter \cite{Ol,Do}.

Let us note that the Jeans mass $M_{J}$ (\ref{71}) is close to the
value $M_{J} \simeq 5 \times 10^{4}\,M_{\odot}$ which was expected
for the instant of recombination with the temperature $T_{dec} =
3800$ K, the redshift $z_{dec} = 1400$ and the mean energy density
in the present-day universe equal to $\rho_{0} = 10^{-29}\
\mbox{g/cm}^{3}$ \cite{ZN}. In the case of isothermal
perturbations in distribution of matter and radiation, when the
radiation is distributed uniformly, while matter is more or less
nonuniform, the Jeans mass for matter was estimated by the value
$M_{J} \simeq 3 \times 10^{5}$ -- $10^{6}\, M_{\odot}$ \cite{Pe}.
This value does not contradict the estimation (\ref{71}) as well.

Among the known (ordinary and hypothetical) particles and fields
we did not find any candidate for dark matter particle with the
mass $\sim 6$ GeV. Since this particle as it is expected should
not participate in any interactions except gravitational, its
registration is highly difficult.

In conclusion we note that the decay scheme (\ref{29}) does not
contradict the quark model of hadrons as for instance the
neutron-proton model of atomic nucleus does not contradict the
fact that the products of decay of radioactive nuclei, as a rule,
are the nuclei of other chemical elements instead of separate
nucleons.

\end{document}